%% file: vuong.tex
\documentclass[doc,apacite]{apa6}\usepackage[]{graphicx}\usepackage[]{color}
\makeatletter
\def\maxwidth{ %
  \ifdim\Gin@nat@width>\linewidth
    \linewidth
  \else
    \Gin@nat@width
  \fi
}
\makeatother

\definecolor{fgcolor}{rgb}{0.345, 0.345, 0.345}

\usepackage{framed}
\makeatletter
 {\par\unskip\endMakeFramed%
 \at@end@of@kframe}
\makeatother

\definecolor{shadecolor}{rgb}{.97, .97, .97}
\definecolor{messagecolor}{rgb}{0, 0, 0}
\definecolor{warningcolor}{rgb}{1, 0, 1}
\definecolor{errorcolor}{rgb}{1, 0, 0}
\newenvironment{knitrout}{}{} 

\usepackage{alltt}
\usepackage{epsfig,amsmath,alltt,bm,url,subcaption}
\usepackage[english]{babel}

\usepackage{tikz}
\usetikzlibrary{decorations.pathmorphing,backgrounds,fit,calc,through,arrows,shapes,trees,positioning,matrix,chains}
\tikzset {crossline/.style={opacity=.15,line width=4.5mm,line cap=round,color=#1},
	method/.style={opacity=.15,fill,rounded corners,color=blue},
	manifest/.style={rectangle,draw,inner sep=0.15em,minimum size=2.5em},
	latent/.style={circle,draw,inner sep=0.15em,minimum size=2.7em},
	error/.style={circle,draw,inner sep=0.15em,minimum size=2em},
	every loop/.style={min distance=1em,in=70,out=110,looseness=3.5},
	highlight/.style={rectangle,rounded corners,fill=red!15,draw,fill opacity=0.5,thick,inner sep=0pt}
}

\usepackage{etoolbox}
\makeatletter
\AtBeginEnvironment{tabular}{%
  \@currsize}%
\makeatother

\title{Testing non-nested structural equation models}
\twoauthors{Edgar C.\ Merkle and Dongjun You}{Kristopher J.\ Preacher}
\twoaffiliations{University of Missouri}{Vanderbilt University}

\abstract{
In this paper, we apply Vuong's (1989) likelihood ratio tests of non-nested
models to the comparison of
non-nested structural equation models.
Similar tests have been previously applied in SEM contexts (especially
to mixture models), though the non-standard output required to conduct
the tests has limited their previous use and study.
We review the theory underlying the tests and show how they can be
used to construct interval estimates for differences in non-nested
information criteria.
Through both simulation and application, we then study the tests'
performance in non-mixture SEMs and describe their general implementation
via free \proglang{R} packages.  The tests offer
researchers a useful tool for
non-nested SEM comparison, with barriers to test implementation now
removed.}

\authornote{This work was supported by National Science
  Foundation grant SES-1061334.  Portions of the work were presented
  at the 2014 Meeting of the Psychometric Society and at the 2015
  International Workshop on Psychometric Computing (Psychoco 2015).
  We thank Achim Zeileis and
  three anonymous reviewers for comments that improved the paper.  All
  remaining errors are solely the responsibility of the authors.
  Correspondence to Edgar C.\ Merkle, Department of
  Psychological Sciences, University of Missouri, Columbia,
  MO 65211.
  Email: {\texttt{merklee@missouri.edu}}.}
\shorttitle{Testing non-nested SEMs}

\let\proglang=\textsf
\let\pkg=\emph

\IfFileExists{upquote.sty}{\usepackage{upquote}}{}
\begin{document}
\maketitle

Researchers frequently rely on model comparisons to test competing
theories.  This is especially true when
structural equation models (SEMs) are used, because the models are
often able to accommodate a large variety of theories.  When competing
theories can be translated into nested SEMs, the
comparison is relatively easy: one can compute likelihood ratio
statistics using the results of the fitted models
\cite<e.g.,>{stesha85}.  The test
associated with this likelihood ratio statistic yields one of two
conclusions: the two models fit equally well, so that the simpler
model is to be preferred, or the more complex model fits better, so
that it is to be preferred.  As is well known, however, the likelihood ratio
statistic does not immediately extend to situations where models are
non-nested.

In the non-nested case, researchers typically rely on information
criteria for model comparison, including the Akaike Information
Criterion \cite<AIC;>{aka74} and the Bayesian Information Criterion
\cite<BIC;>{sch78}.  One computes an
AIC or BIC for the two models, then selects the model with the lowest
criterion as ``best.''  Thus, the applied conclusion differs slightly
from the likelihood ratio test (LRT): we conclude from the information
criteria that one or the other model is better, while we conclude from
the LRT either that the complex model is better or that there is
insufficient evidence to differentiate between model fits.

While information criteria can be applied to non-nested models, the
popular ``select the model with the lowest'' decision criterion can be
problematic.  In particular, \citeA{premer12}
showed that BIC exhibits large variability at the sample sizes
typically used in SEM contexts.  Thus, the model that is preferred
for a given sample often will not be preferred in new samples.
Preacher and Merkle studied a series of nonparametric bootstrap
procedures to
estimate sampling variability in BIC, but no procedure succeeded in
fully characterizing this variability.

A problem with the ``select the model with the lowest'' decision
criterion
involves the fact that one can never conclude that the models fit equally.  There may often be situations where the models exhibit
``close'' values of the information criteria, yet one of the models is
still selected as best.  To handle this issue,
\citeA{porwu13} developed a parametric bootstrap
method that allows one to conclude that the two models are equally
good (in addition to concluding that one or the other model fits 
better).  Their results indicated that the procedure is promising,
though it is also computationally expensive: one must draw
a large number of bootstrap samples from each of the two fitted models,
then refit each model to each bootstrap sample.

In this paper, we study formal tests of non-nested models that allow
us to conclude that one model fits better than the other, that the two
models exhibit equal fit, or that the two models are
indistinguishable in the population of interest.  The tests are based on the theory of Vuong \citeyear{vuo89},
and one of the tests is popularly applied to the comparison of mixture
models with
different numbers of components \cite<including count regression
models and factor mixture models;>{gre94,lomen01,nylmut07}.  
While some authors have recently described problems with mixture model applications \cite{jef03,wil15}, the tests have the
potential to be very useful in general SEM contexts.  This is because
non-nested SEMs are commonly observed throughout psychology \cite<e.g.,>{fregar07,kim02,santod02}.

\citeA{levhan07,levhan11} have previously studied the
application of Vuong's \citeyear{vuo89} theory to structural equation
models, describing relevant background and proposing steps by which
researchers can carry out tests of non-nested models.  Levy and
Hancock bypass an important step of the theory due to the
non-standard model output required, instead requiring researchers to
algebraically examine the candidate models and to potentially carry
out likelihood ratio tests between each candidate model and a constrained
version of the models.  This procedure can accomplish
the desired goal, but it also requires a considerable amount of analytic
and computational work on the part of the user.  We instead study the
tests as Vuong originally proposed them, using the non-standard model
output that is required.
This study is aided by our general implementation of the tests, available via the R package \pkg{nonnest2} \cite{nonnest2}.

In the following pages, we first describe the relevant theoretical
results from Vuong \citeyear{vuo89}.  We also show how the theory can
be used to obtain confidence intervals for differences in BICs (and
other information criteria) associated with non-nested models.
Next, we apply the tests to data on teacher burnout, which were originally
examined by Byrne \citeyear{byrne1994burnout}.
Next, we describe the results of three simulations
that illustrate test properties in the context of
SEM.  Finally, we discuss recommendations, extensions, and practical issues.

\section{Theoretical Background}

In this section, we provide an overview of the theory
underlying the test statistics.  The overview is largely based on
Vuong \citeyear{vuo89}, and the reader is referred to that paper for
further detail.  For alternative overviews of the theory, see
\citeA{gol00} and \citeA{levhan07}.  The theory is applicable to many
models estimated via Maximum Likelihood (ML), though we focus on
structural equation models here.

We consider situations where two candidate structural
equation models, $M_A$ and
$M_B$, are to be compared using a dataset $\bm{X}$ with $n$ cases and
$p$ manifest variables.  $M_A$ may be
represented by the equations
\begin{align}
    \bm{x}_i &= \bm{\nu}_A + \bm{\Lambda}_A \bm{\eta}_{A,i} +
    \bm{\epsilon}_{A,i} \\
    \bm{\eta}_{A,i} &= \bm{\alpha}_A + \bm{B}_A\bm{\eta}_{A,i} + \bm{\zeta}_{A,i},
\end{align}
where $\bm{\eta}_{A,i}$ is a vector containing the latent
variables in $M_A$; $\bm{\epsilon}_{A,i}$ and $\bm{\zeta}_{A,i}$ are
zero-centered residual vectors, independent across values of $i$; $\bm{\Lambda}_A$ is a
matrix of factor loadings; and
$\bm{B}_A$ contains parameters
that reflect directed paths between latent variables.  The second
model, $M_B$, is defined similarly, and we restrict ourselves to
situations where the residuals and latent variables are assumed to follow
multivariate normal
distributions (though this distributional assumption is not required
to use the test statistics; see the General Discussion).

The $M_A$ parameter vector, $\bm{\theta}_A$, includes the parameters
in $\bm{\nu}_A$, $\bm{\Lambda}_A$, $\bm{\alpha}_A$, and
$\bm{B}_A$, along with variance and covariance parameters related
to the latent variables and residuals.  These parameters imply a marginal
multivariate normal distribution for $\bm{x}_i$ with a specific mean
vector (typically $\bm{\nu}_A$) and covariance matrix \cite<see,
e.g.,>{broarm95}, which allows us to estimate the model via ML.  In
particular, we will choose $\hat{\bm{\theta}}_A$ to maximize the
log-likelihood ($\ell()$):
\begin{equation*}
  \ell(\bm{\theta}_A; \bm{x}_1, \dots, \bm{x}_n) ~=~
    \sum_{i = 1}^n \ell(\bm{\theta}_A; \bm{x}_i) ~=~
    \sum_{i = 1}^n  \log f_A(\bm{x}_i; \bm{\theta}_A),
\end{equation*}
where $f_A(\bm{x}_i; \bm{\theta}_A)$ is the probability density function (pdf) of the multivariate
normal distribution, with mean vector and covariance matrix implied by
$M_A$ and its parameter vector $\bm{\theta}_A$.  Similarly, the parameters $\hat{\bm{\theta}}_B$ are chosen to maximize the log-likelihood $\ell(\bm{\theta}_B; \bm{x}_1, \dots,
\bm{x}_n)$.

Instead of defining the ML estimates as above, we could equivalently define them via the gradients
\begin{align*}
  s(\hat{\bm{\theta}}_A; \bm{x}_1, \dots,
\bm{x}_n) = \sum_{i=1}^{n} {\bm s}(\hat{\bm{\theta}}_A; \bm{x}_i) &~=~ 0 \\
  s(\hat{\bm{\theta}}_B; \bm{x}_1, \dots,
\bm{x}_n) = \sum_{i=1}^{n} {\bm s}(\hat{\bm{\theta}}_B; \bm{x}_i) &~=~ 0,
\end{align*}
where the gradients sum {\em scores} across individuals (where scores are defined as the casewise contributions to the gradient).  The above equations simply state that we choose parameters such that the gradient (first derivative of the likelihood function) equals zero.  Assuming that $M_A$
has $k$ free parameters, the associated score function may be explicitly defined as
\begin{equation}
  \label{eq:score}
  {\bm s}({\bm \theta}_A; \bm{x}_i) ~=~ \left(
    \frac{\partial \ell({\bm \theta}_A; \bm{x}_i)}{\partial \theta_{A,1}},
    \dots,
    \frac{\partial \ell({\bm \theta}_A; \bm{x}_i)}{\partial \theta_{A,k}}
  \right)^\prime,
\end{equation}
with the score function for $M_B$ defined similarly (where the number
of free parameters for $M_B$ is $q$ instead of $k$).
We also define $M_A$'s expected information matrix as
\begin{equation}
    \label{eq:infmat}
    \bm{\mathrm{I}}(\bm{\theta}_A) = -\mathrm{E}\ \frac{\partial^2
      \ell({\bm \theta}_A; \bm{x}_1, \ldots, \bm{x}_n)}{\partial
      \bm{\theta}_{A} \partial \bm{\theta}_A^\prime}
\end{equation}
where, again, the information matrix of $M_B$ is defined similarly.

\begin{figure}
  \caption{Path diagram reflecting the models used in the
    simulation. Model A is the data-generating model, with the loading
  labeled `A' varying across conditions.}
  \label{fig:pathdiag1}
  \centering
  \resizebox{0.6\textwidth}{!}{
    \input{simDiagram}
  }
\end{figure}
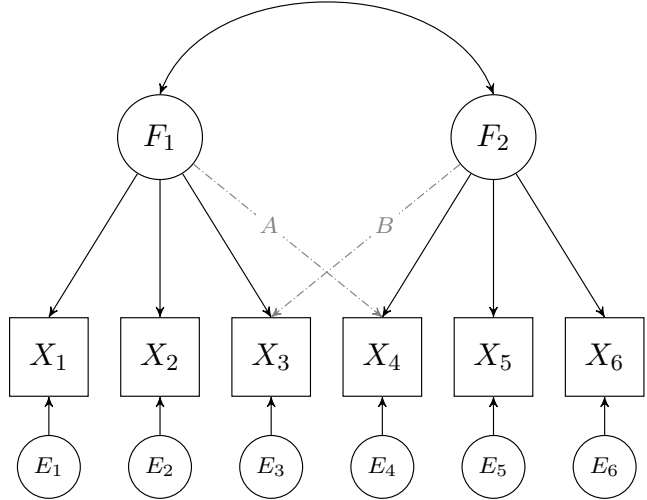

The statistics described here can be used in general model comparison
situations, where one is interested in which of two candidate models
($M_A$ and $M_B$) is closest to the data-generating model in
Kullback--Leibler distance \cite{kullei51}.  Letting $g()$ be the
density of the data-generating model (which is generally unknown), the
distances for $M_A$ and $M_B$ can be
denoted $\text{KL}_{Ag}$ and $\text{KL}_{Bg}$, respectively.
The distance $\text{KL}_{Ag}$ may be explicitly written as
\begin{align}
    \label{eq:kl}
    \text{KL}_{Ag} &= \displaystyle \int \log \left ( \frac{g(\bm{x})}{f_A(\bm{x};
          \bm{\theta}^*_A)} \right ) g(\bm{x}) d\bm{x} \\
    \label{eq:kl2}
    &= \mathrm{E} \left [ \log(g(\bm{x})) \right ] - \mathrm{E} \left [\log(f_A(\bm{x};
        \bm{\theta}^*_A)) \right ],
\end{align}
where $\bm{\theta}^*_A$ is the $M_A$ parameter vector that minimizes
this distance (also called the ``pseudo-true'' parameter vector) and
where the expected values are taken with respect to $g()$.  The
``pseudo-true'' name arises from the fact that $\bm{\theta}^*_A$
usually does not reflect the true parameter vector (because the
candidate model $M_A$ is usually incorrect).  However, the parameter
vector is pseudo-true in the sense that it allows $M_A$ to most
closely approximate the truth.

\subsection{Relationships Between Models}
Relationships between pairs of candidate models may be characterized
in multiple manners.  Familiarly, {\em nested} models are those for which
the parameter space associated with one model is a subset of the
parameter space associated with the other model; every set of
parameters from the less-complex model can be translated into an equivalent set of
parameters from the more-complex model.  Similarly, {\em non-nested}
models are those whose parameter spaces each include some unique points.
Aside from these two broad, familiar classifications, however, we may
define other relationships between models.  These include the concepts
of {\em equivalence}, {\em overlappingness}, and {\em strict
  non-nesting}.  The latter two concepts refer to specific types of
non-nested models.  All three are described below.

Many SEM researchers are familiar with the concept of model equivalence \cite<e.g.,>{hermar13,leeher90,macweg93}:
two seemingly-different SEMs yield exactly the same implied moments
(mean vectors and covariance matrices) and fit statistics when fit to any dataset.
SEM researchers are less familiar with the concept of overlapping
models: these are models that yield exactly the same implied moments
and fit statistics for some populations but {\em not}
for others.  Further, even if the models yield identical moments in
our population of interest, they will be exactly identical only in the
population.  Fits to sample data will generally yield different
moments and fit statistics.

An example of overlapping models is displayed in
Figure~\ref{fig:pathdiag1}.  Both $M_A$ and $M_B$ are two-factor
models, where $M_A$ has a
free path from $\eta_1$ to $X_4$ and $M_B$ has a free path
from $\eta_2$ to $X_3$.  These models are overlapping: their predictions
and fit statistics will usually differ, but they will be the same in
populations where
the parameters labeled `A' and `B' both equal zero.  Even if these two
paths equal zero in the population, they generally will not equal zero
when the models are fit to sample data.  Therefore, the models will
not have the same implied moments and fit statistics when fit to sample
data.  This means that, given sample data, we must test whether or not the models are {\em
  distinguishable} in the population of interest.  Note that
{\em overlappingness} is a general relationship between two models
regardless of the focal population, whereas
{\em distinguishability} is a specific relationship between models in
the context of a single focal population.

Overlapping models are one sub-classification of non-nested models.  The other
sub-classification is {\em
  strictly non-nested} models; these are models whose parameter spaces
do not overlap at all.  In other words, strictly non-nested models
never yield the same implied moments and fit statistics for any
population.  Strictly non-nested models may have different functional
forms (say, an exponential growth model versus a logistic growth model) or may make
different distributional assumptions.

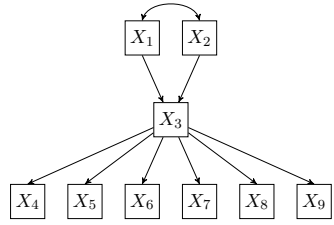
\begin{figure}%
  \caption{Path diagrams reflecting the models used in Simulation
    2.} \label{fig:sim2}
  \centering
  \begin{subfigure}[t]{0.4\textwidth}%
    \centering
    \captionsetup{justification=centering,font=scriptsize}
    \resizebox{0.77\textwidth}{!}{
      \input{modelA}
    }
    \caption*{Model A ($df=27$)}\label{fig:sim2A}
  \end{subfigure}%
  ~
  \begin{subfigure}[t]{0.4\textwidth}%
    \centering
    \captionsetup{justification=centering,font=scriptsize}
    \resizebox{0.97\textwidth}{!}{
      \input{modelB}
    }
    \caption*{Model B ($df=21$)}\label{fig:sim2B}
  \end{subfigure}%

  \begin{subfigure}[t]{0.4\textwidth}%
    \centering
    \captionsetup{justification=centering,font=scriptsize}
    \resizebox{0.75\textwidth}{!}{
      \input{modelC}
    }
    \caption*{Model C ($df=20$)} \label{fig:sim2C}
  \end{subfigure}%
  ~
  \begin{subfigure}[t]{0.4\textwidth}%
    \centering
    \captionsetup{justification=centering,font=scriptsize}
    \resizebox{0.75\textwidth}{!}{
      \input{modelD}
    }
    \caption*{Model D ($df=26$)} \label{fig:sim2D}
  \end{subfigure}%
\end{figure}

It is often difficult to know whether or not candidate SEMs are overlapping.  Consider the path models in Figure~\ref{fig:sim2}, which reflect four potential hypotheses about the relationships between nine observed variables.  These models are obviously non-nested, but are they overlapping or strictly non-nested?  Assuming that all models employ
the same data distribution (typically multivariate normal), then the models will be indistinguishable in populations where all
observed variables are independent of one another.  Therefore, SEMs
that employ the same form of data distribution will typically be
overlapping.  In other situations, it may be difficult to tell whether
or not the candidate models are overlapping.
Thus, it is
important to test for distinguishability when doing model comparisons:
if the observed data imply that the models are indistinguishable in
the population of interest, then
there is no point in further model comparison.  This is especially
relevant to the use of information criteria for non-nested model comparison, where one is guaranteed to select a candidate model as better (at
least, using the standard decision criteria).
Additionally, as we will see below, when a pair of models is
non-nested, the limiting 
distribution of the likelihood ratio statistic depends on whether or not the
models are distinguishable.

The above discussion suggests a sequence of tests for comparing two
models.  Assuming that the models are not equivalent to one
another (regardless of population), we must
establish that the models are
distinguishable in the population of interest.  Assuming that
the models are distinguishable, we can then compare the models' fits and
potentially select one as better.  Below, we describe test statistics
that can be used in this sequence.

\subsection{Test Statistics}
Vuong's \citeyear{vuo89} tests of distinguishability and of model fit
utilize the terms $\ell(\hat{\bm{\theta}}_A; \bm{x}_i)$
and $\ell(\hat{\bm{\theta}}_B; \bm{x}_i)$ for $i=1,\ldots,n$, which
are the casewise likelihoods evaluated at the ML estimates.
For the purpose of SEM, we focus on two separate statistics
that Vuong proposed.  One statistic tests whether or not models are
distinguishable, and the other tests the fit of non-nested,
distinguishable models.  These tests proceed sequentially in
situations where we are unsure whether or not the two models are
distinguishable in the population of interest.  If we know in advance that two models are not overlapping, we can proceed directly to the second test.

To gain an intuitive feel for the tests,
imagine that we fit both $M_A$ and $M_B$ to a data set and then obtain
$\ell(\hat{\bm{\theta}}_A; \bm{x}_i)$ and $\ell(\hat{\bm{\theta}}_B;
\bm{x}_i)$ for $i=1,\ldots,n$.
To test whether or not
the models are distinguishable, we can calculate a likelihood ratio for each case
$i$ and examine the variability in these $n$ ratios.
If the models are indistinguishable, then these ratios should be similar for all
individuals, so that their variability is close to zero.
If the models are distinguishable, then the variability in the casewise
likelihood ratios characterizes general sampling variability in the likelihood ratio
between $M_A$ and $M_B$, allowing for a formal model comparison test
that does not require the models to be nested.

To formalize the ideas in the previous paragraph, we characterize the
population variance in individual likelihood ratios of $M_A$ vs $M_B$ as
\begin{equation*}
    \omega^2_* = \text{var} \left [ \log \frac{f_A(\bm{x}_i; \bm{\theta}^*_A)}{
          f_B(\bm{x}_i; \bm{\theta}^*_B)} \right ]\ \ i=1,\ldots,n,
\end{equation*}
where this variance is taken with respect to the pseudo-true parameter
vectors introduced in Equation~\eqref{eq:kl}.  This means that we make
no assumptions that either candidate model is the true model.
Using the above equation, hypotheses for a test of model distinguishability
may then be written as
\begin{align}
    \label{eq:h0om2}
    H_0\colon&\ \omega^2_* = 0\\
    \label{eq:h1om2}
    H_1\colon&\ \omega^2_* > 0,
\end{align}
with a sample estimate of $\omega^2_*$ being
\begin{equation}
    \label{eq:sampom}
    \hat{\omega}^2_* = \frac{1}{n} \displaystyle \sum_{i=1}^n \left [
        \log \frac{f_A(\bm{x}_i; \hat{\bm{\theta}}_A)}{
          f_B(\bm{x}_i; \hat{\bm{\theta}}_B)} \right ]^2 - \left [
        \frac{1}{n} \displaystyle \sum_{i=1}^n  \log
        \frac{f_A(\bm{x}_i; \hat{\bm{\theta}}_A)}{
          f_B(\bm{x}_i; \hat{\bm{\theta}}_B)} \right ]^2.
\end{equation}
Division by $(n-1)$ instead of $n$ is also possible to reduce bias in
the estimate, though this will have little impact at the sample sizes
typically observed in SEM applications.

Vuong shows that, under~\eqref{eq:h0om2} and mild regularity
conditions (ensuring that second derivatives of the likelihood function exist, observations are i.i.d., and the ML estimates are unique and not on the boundary), $n \hat{\omega}^2_*$ is asymptotically
distributed as a
particular weighted sum of $\chi^2$ distributions.  Weighted sums of
$\chi^2$ distributions arise when we sum the squares of
normally-distributed variables; the normally-distributed variables
involved in the test statistics here are the ML estimates
$\hat{\bm{\theta}}_A$ and 
$\hat{\bm{\theta}}_B$.  The weights involved in this sum are obtained
via
the {\em squared} eigenvalues of a matrix $\bm{W}$ that arises from the candidate models'
scores (Equation~\eqref{eq:score}) and
information matrices (Equation~\eqref{eq:infmat}); see the Appendix for details.  This result
immediately allows us to test~\eqref{eq:h0om2} using results from the
two fitted models.  If the null hypothesis is not rejected, we
conclude that the models cannot be distinguished in the population of
interest.  In the case where models are nested, this conclusion
would lead us to prefer the model with fewer degrees of freedom.  In
the case where models are not nested, the two candidate
models may have the same degrees of freedom.  Thus, depending on the
specific models being compared, we might not prefer either model.

The software requirements for carrying out the test of~\eqref{eq:h0om2} is
somewhat non-standard.  For each candidate model, we need to obtain
the scores from~\eqref{eq:score} and information matrix
from~\eqref{eq:infmat}.  We then need to arrange this output in
matrices, do some multiplications to obtain a new matrix, and obtain
eigenvalues of this new 
matrix.  Further, we need the ability to evaluate quantiles of 
weighted sum of $\chi^2$ distributions.
The difficulty in obtaining these results led
\citeA{levhan07} to bypass the test of model distinguishability and instead
conduct an algebraic model comparison that provides evidence about whether or not the
models are distinguishable.
Whereas this algebraic comparison is
reasonable, it does not always indicate whether or not models are distinguishable.  Further, it is more complicated for the applied researcher who wishes
to use these tests (assuming that an implementation of the statistical
test is available).


Assuming that the null hypothesis from~\eqref{eq:h0om2} is rejected
(i.e., that the models are distinguishable), we may compare the models via a
non-nested LRT.  We can write the
hypotheses associated with this test as
\begin{align}
    \label{eq:h0lrt}
    H_0\colon&\ \mathrm{E}[\ell(\hat{\bm{\theta}}_A; \bm{x}_i)] =
    \mathrm{E}[\ell(\hat{\bm{\theta}}_B; \bm{x}_i)] \\
    \label{eq:h1alrt}
    H_{1A}\colon&\ \mathrm{E}[\ell(\hat{\bm{\theta}}_A; \bm{x}_i)] >
    \mathrm{E}[\ell(\hat{\bm{\theta}}_B; \bm{x}_i)] \\
    \label{eq:h1blrt}
    H_{1B}\colon&\ \mathrm{E}[\ell(\hat{\bm{\theta}}_A; \bm{x}_i)] <
    \mathrm{E}[\ell(\hat{\bm{\theta}}_B; \bm{x}_i)],
\end{align}
where the expectations arise from the K-L distance in
Equation~\eqref{eq:kl2} (note that we don't need to consider the
expected value of $\log(g(\bm{x}))$
because it is constant across candidate models).  The hypothesis
$H_0$ above states that the K-L distance between $M_A$ and the truth
equals the K-L distance between $M_B$ and the truth; this is like
stating that the two models have equal population discrepancies.
The test is
written to be directional, so that one chooses either $H_{1A}$ or
$H_{1B}$ prior to carrying out the test.  In practice, however, a
two-tailed test is often carried out, with a single model being
preferred in the situation where $H_0$ is rejected.  This follows the
framework of \citeA{jontuk01}, whereby there are three possible
conclusions available to researchers: $M_A$ is closer to the truth (in
K-L distance) than $M_B$; $M_B$ is closer to the truth than $M_A$; or there is
insufficient evidence to conclude that either model is closer to the truth
than the other.

For non-nested, distinguishable models, Vuong shows that
\begin{equation}
    \label{eq:zlrt}
    \text{LR}_{AB} = n^{-1/2} \displaystyle \sum_{i=1}^n \log \frac{f_A(\bm{x}_i;
      \hat{\bm{\theta}}_A)}{f_B(\bm{x}_i; \hat{\bm{\theta}}_B)}
    ~\overset{d}{\rightarrow}~ N(0, \omega^2_*)
\end{equation}
under~\eqref{eq:h0lrt} and the regularity conditions noted above.  Thus, we
obtain critical values and $p$-values by comparing the
non-nested LRT statistic to the standard normal distribution.
Assuming that the desired Type I error rate is $\alpha_2$ for this
test and that the desired Type I error rate is $\alpha_1$ for the test
of~\eqref{eq:h0om2}, Vuong shows that the sequence of tests has an
overall Type I error rate that is bounded from above by
$\text{max}(\alpha_1, \alpha_2)$.  In practical applications, it is
customary to set $\alpha_1 = \alpha_2$.

Assuming that the null hypothesis from~\eqref{eq:h0om2} is not
rejected (i.e., that the models are indistinguishable), Vuong shows
that $2n^{1/2}\text{LR}_{AB}$ follows a weighted sum of $\chi^2$
distributions, where the weights are obtained from the {\em unsquared}
eigenvalues of the same matrix $\bm{W}$ that arose in the test
of~\eqref{eq:h0om2}.  This result is not typically used in the case of
non-nested models: we first need the test of $\omega^2_*$ to determine
the limiting distribution that we should use for the likelihood ratio
(either the normal distribution from~\eqref{eq:zlrt} or the weighted
sum of $\chi^2$ distribution described here).
If the test of $\omega^2_*$ indicates
indistinguishable models, however, then there is no point in further testing the
models via $2n^{1/2}\text{LR}_{AB}$.  If the test of $\omega^2_*$ indicates distinguishable
models, then we rely on the limiting distribution
from~\eqref{eq:zlrt} instead of the weighted sum of $\chi^2$s.  The
result described in this paragraph can be used to compare nested
models, however, and we return to this topic in the next subsection.

The test statistics described above are
implemented for general multivariate normal SEMs (and other
models) in the free \proglang{R} package \pkg{nonnest2}
\cite{nonnest2}.  Models are first estimated via \pkg{lavaan}
\cite{lavaan11}, then \pkg{nonnest2} computes the test statistics
based on the fitted models' output.
In the following sections, we describe
ways in which these ideas can be extended to test nested models and
to test information criteria.

\subsection{Testing Nested Models} 
In the situation where $M_B$ is nested within $M_A$, the
likelihood ratio and the variance statistic $n\omega_*^2$ can each be used
to construct a unique test of $M_A$ versus $M_B$. For nested models,
the null hypotheses~\eqref{eq:h0om2} and~\eqref{eq:h0lrt} can be shown to be
the same as the traditional null hypothesis:
\begin{align}
    \label{eq:tradh0}
    H_0\colon&\ \bm{\theta}_A \in\ h(\bm{\theta}_B) \\
    \label{eq:tradh1}
    H_1\colon&\ \bm{\theta}_A \not\in\ h(\bm{\theta}_B),
\end{align}
where $h()$ is a function translating the $M_B$ parameter vector
to an equivalent $M_A$ parameter vector.
In the general
case, where $M_A$ is not assumed to be correctly specified, the
statistics $n\hat{\omega}^2_*$ and $2n^{1/2}\text{LR}_{AB}$
(see Equations~\eqref{eq:sampom} and~\eqref{eq:zlrt}) both
strongly converge \cite<i.e., almost surely;>{casber02} to weighted sums of $\chi^2$
distributions under the null hypothesis from~\eqref{eq:tradh0}.
The specific weights differ between the two statistics; the weights
associated with $n\hat{\omega}^2_*$ are the squared eigenvalues
of a $\bm{W}$ matrix that is defined in the Appendix, whereas the weights
associated with $2n^{1/2}\text{LR}_{AB}$ are the unsquared
eigenvalues of the same $\bm{W}$ matrix.
This result differs from the usual multivariate normal SEM
derivations \cite<e.g.,>{ameand90,stesha85}, which employ either an
assumption that $M_A$
is correctly specified or that the population parameters drift toward
a point that is contained in $M_A$'s parameter space
\cite<see>[for further discussion of this point and fit assessment of single models]{chusha09}.
Under the
assumption that $M_A$ is correctly specified, the statistics
$n\hat{\omega}^2_*$ and $2n^{1/2}\text{LR}_{AB}$
weakly converge (i.e., in distribution) to the usual $\chi^2_{\text{df}=q-k}$ distribution
under~\eqref{eq:tradh0}.  Hence, the framework here provides a more
general characterization of the nested LRT than do traditional
derivations.

\subsection{Testing Information Criteria}
Model selection with AIC
or BIC (i.e., selecting the model with the lowest) involves adjustment
of the
likelihood ratio by a constant term that penalizes the two models for
complexity.  Thus, as Vuong \citeyear{vuo89} originally described, the
above results can be extended to test differences in AIC
or BIC.
To show this formally, we focus on BIC and write the BIC
difference between two models as:
\begin{equation*}
    \label{eq:bicdiff}
    \text{BIC}_A - \text{BIC}_B = (k \log n - q \log n) -  2
    \displaystyle \sum_{i=1}^n \log \frac{f_A(\bm{x}_i;
      \hat{\bm{\theta}}_A)}{f_B(\bm{x}_i; \hat{\bm{\theta}}_B)},
\end{equation*}
where $k$ and $q$ are the number of free parameters for $M_A$ and
$M_B$, respectively.  This shows that we are simply taking a linear
transformation of the usual likelihood ratio, so that the
test of~\eqref{eq:h0om2} and the result from~\eqref{eq:zlrt} apply
here.  In particular, if models are
indistinguishable, then one could select the model that BIC
penalizes the least (i.e., the model with fewer parameters).
If models are distinguishable, then we may formulate a hypothesis that
$\text{BIC}_A = \text{BIC}_B$.  Under this hypothesis, the result
from~\eqref{eq:zlrt} can be used to show that
\begin{equation*}
    n^{-1/2} \left [ ((k-q)\log n) - 2\displaystyle \sum_{i=1}^n \log
      \frac{f_A(\bm{x}_i;
      \hat{\bm{\theta}}_A)}{f_B(\bm{x}_i; \hat{\bm{\theta}}_B)} \right ]
    ~\overset{d}{\rightarrow}~ N(0, 4 \omega^2_*).
\end{equation*}
This result can be used to obtain an ``adjusted'' test statistic that
accounts for the models' relative complexity.  Alternatively, we
prefer to use this result to construct a
$100\times (1-\alpha)$\% confidence interval associated with the BIC
difference.  This is obtained via
\begin{equation}
    \label{eq:bicdiffint}
    (\text{BIC}_A - \text{BIC}_B) \pm z_{1-\alpha/2} \sqrt{4n\omega^2_*},
\end{equation}
where $z_{1-\alpha/2}$ is the variate at which the cumulative
distribution function (cdf) of the
standard normal distribution equals $(1 - \alpha/2)$.
To coincide with the tests described previously, $\alpha$ should be
the same as the $\alpha$ level used for the test of distinguishability.

To our knowledge, this is the first analytic confidence interval for
a non-nested difference in information criteria that has been
presented in the SEM literature.
This confidence interval is simpler to calculate than bootstrap
intervals \cite<see, e.g.,>[for a discussion of bootstrap
procedures]{premer12}, and, as shown later, its coverage is often comparable.
The bootstrap intervals may still be advantageous if regularity conditions are violated.


\subsection{Relation to the Nesting and Equivalence Test}
\citeA{bentler2010testing} describe a Nesting and Equivalence Test
(NET) that assesses whether two models provide exactly the same fit to
sample data, relying on the fact that equivalent models can perfectly
fit one another's implied mean vectors and covariance matrices.  As
previously described, this differs from the ``indistinguishability''
characteristic that is relevant to the tests in this paper.
Indistinguishable models provide exactly the same fit in the
population but not necessarily to sample data.




The NET procedure is convenient and computationally simple, and it
is generally suited to examining whether two models are globally nested or
equivalent across large sets of covariance matrices
\cite<though see>[for some pathological cases]{bentler2010testing}.
It cannot, however, inform us about whether or not two models are
distinguishable based on the sample data.  In other words, equivalent models are indistinguishable, but indistinguishable models are not necessarily equivalent.
Thus, the two methods are complementary: we can use the NET to determine
whether two models can possibly be distinguished from each another,
while we can use the test of~\eqref{eq:h0om2} to determine whether two
models can be distinguished in the population of interest. 
In the following sections, we study the tests' applications to SEM using
both simulation and real data.

\section{Application: Teacher Burnout}

\subsection{Background}

\citeA{byrne1994burnout} tested the impact of organizational
and personality variables on three dimensions of teacher burnout. The application here is 
limited to the sample of elementary teachers only ($n = 599$; intermediate and secondary school teachers were also observed)
and utilizes models that are related to those presented in Chapter~6 of
 \citeA{byrne2009structural}.

%
\begin{figure}
  \caption{Path diagram of candidate Models 1, 2, and 3, which are
    related to those originally specified by Byrne (2009). Only latent
    variables are displayed, and covariances between exogenous
    variables are always estimated.}
  \centering
  \resizebox{0.6\textwidth}{!}{
    \input{burnoutDiagram}
  }
  \label{fig:burnoutDiagram}
\end{figure}
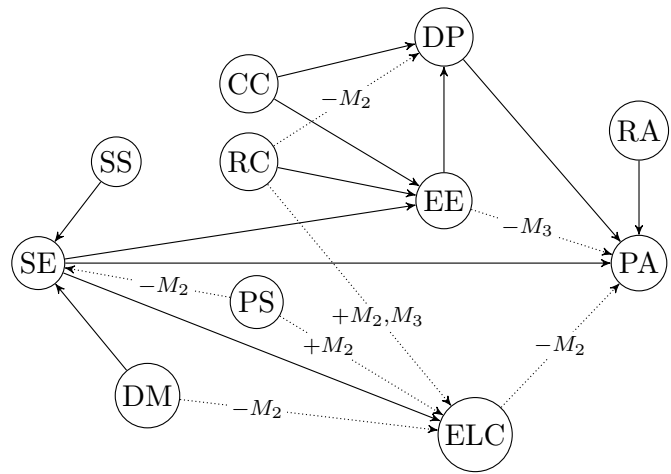
%


%

\subsection{Method}

The candidate models that we consider are illustrated in
Figure~\ref{fig:burnoutDiagram}, which shows only the latent variables
included in the model (and not the indicators of the latent variables
or variance parameters).
The figure displays candidate model $M_1$, with dotted lines reflecting
parameters that are added or removed in models $M_2$ and $M_3$.
Each pair of models is non-nested, so BIC (or some other information
criterion) would typically be used to select a model from the set.
Alternatively, we can use the statistics described in this paper
to study the models' distinguishability and fit in greater detail.


%
To expand on the model comparison procedure, we first use the NET
\cite{bentler2010testing} to
determine whether or not pairs of models are equivalent to one another.  For
models that are not equivalent, we then use Vuong's
distinguishability test to make inferences about whether or not each
pair of models is distinguishable based on the focal population of elementary teachers.  Finally,
we use Vuong's non-nested LRT to study the candidate models' relative
fit.  If desired, the latter test can be accompanied by BIC
statistics and interval estimates of BIC differences.

\subsection{Results}

To mimic a traditional comparison of non-nested models, we first
examine the three candidate models' BICs.  We find that BIC decreases
as we move from $M_1$ to $M_3$ ($\text{BIC}_1$ =
40040.7; $\text{BIC}_2 =$
39994.1; $\text{BIC}_3 =$
39978.9), which would lead
us to prefer $M_3$.  Additionally, the BIC difference between $M_3$
and its closest competitor, $M_2$, is about 15.  Using the
\citeA{raf95} ``grades of evidence''
for BIC differences, we would
conclude that there is ``very strong'' evidence for $M_3$ over the
other models.

We now undertake a larger model comparison via the NET and the Vuong
tests, comparing each candidate model to $M_3$.  In applying the NET procedure,
we find that neither $M_1$ nor $M_2$ is equivalent to $M_3$
($\hat{F}_{13} =$ 57.9;
 $\hat{F}_{23} =$ 13.4).
Next, we test whether or not each pair of models is distinguishable,
using the test of~\eqref{eq:h0om2}.  We find that both $M_1$ and $M_2$
are distinguishable from $M_3$ ($\hat{\omega}^2_{13} =$
0.12, $p < .01$; 
$\hat{\omega}^2_{23} =$
0.05, $p < .01$).
This yields
evidence that the models can be distinguished from one another based
on the population of interest, so
that it makes sense to further compare the models' fits.

To compare model fits, we use Vuong's non-nested LRT of~\eqref{eq:h0lrt}.
We find that $M_3$ fits better than $M_1$ ($z_{13} =$
2.86, $p =$
0.002), reinforcing
the BIC results described above.  The test of $M_2$ vs $M_3$ differs
from the BIC results, however.  Here, we do not reject the hypothesis that the
two models' fits are equal in the population ($z_{23} =$
0.84, $p=$
0.20).
Furthermore, the 95\% confidence interval associated with the BIC
difference of $M_2$ vs.\ $M_3$
is $(-5.3,
35.7)$.  This
overlaps with zero, providing evidence that we cannot prefer either
model after adjusting for differences in model complexity.

The above results imply that, despite having ``very strong'' evidence
for $M_3$ via traditional BIC comparisons, the fits of $M_2$ and $M_3$ are
sufficiently close that we cannot prefer either model over the other.
Vuong's methodology provided us the ability to draw these conclusions in
a straightforward manner that can be generally applied to SEM.
A reviewer noted that, for the example here, similar conclusions could
be drawn by specifying a larger model within which all candidate
models are nested.  This
larger model would include free parameters associated with every path
(solid or dotted) in Figure~\ref{fig:burnoutDiagram}, and we could
compare each candidate model to this larger model via traditional
likelihood ratio tests.  This strategy is useful, though it is not as
general as the Vuong methodology.  For example, the ``larger model''
strategy could not be employed if we compared models with differing
distributional assumptions.
Further, estimation of the larger model may be difficult or impossible
in some situations (resulting in, e.g., the estimation algorithm
failing to converge to the ML estimates).

In the next sections, we further study the tests' abilities via
simulation.

\section{Simulation 1: Overlapping Models}

In Simulation 1, our data-generating model is sometimes a special case
of both candidate models, so that the models are sometimes indistinguishable.   We study the test of~\eqref{eq:h0om2}'s ability to pick up the indistinguishable models, and
we also study the non-nested LRT's (of~\eqref{eq:h0lrt}) ability to compare
models that are judged to be distinguishable.  Finally, we compare the results obtained with these two novel tests to the use of (i) the \citeA{bentler2010testing} NET and (ii) the BIC for model comparison.
While the true model is included in the set of estimated models for simplicity, this need not be assumed for Vuong's tests to be valid.

\subsection{Method}
The two candidate models are displayed in the previously-discussed
Figure~\ref{fig:pathdiag1}:
both are two-factor models, and they differ in which loadings are
estimated.  The data-generating model, $M_A$, has an extra loading
from the first factor to the fourth indicator (labeled `A' in the
figure).  The second model, $M_B$, instead has an extra loading from
the second factor to the third indicator (labeled `B' in the figure).

To study the tests described in this paper, we set the data-generating
model's parameter values equal to the parameter estimates obtained
from a two-factor model fit to the \citeA{holswi39} data
(using the scales that load on the ``textual'' and ``speed''
factors).  Additionally, we manipulated the
magnitude of the `A' loading during data generation: this loading
could take values of $d= 0, .1, \ldots, .5$.  In the condition where
$d=0$, the data-generating model is a special case of both candidate
models.  In other conditions, $M_A$ is preferable to $M_B$.  However,
when $d$ is close to zero, the tests may still indicate that the models are indistinguishable from one another.

Simulation conditions were defined by $d= 0,
.1, \ldots, .5$ and by $n=200, 500,$ and 1,000.  In each condition,
we generated 3,000 datasets and fit both $M_A$ and $M_B$ to the
data.  We then computed five statistics: the NET
\cite{bentler2010testing}; the distinguishability test
of~\eqref{eq:h0om2}; the LRT of~\eqref{eq:h0lrt}; and each model's BIC.
The BIC is not required here (the Vuong tests can be used in place of
the BIC), but we included it for comparison.  Using each
statistic, we recorded whether or not $M_A$ was
favored for each dataset.  To be specific, we counted each statistic as
favoring $M_A$ if: (i) the NET implied that models were not
equivalent, (ii) the test of~\eqref{eq:h0om2} implied that models were
distinguishable at $\alpha=.05$, (iii) the test of~\eqref{eq:h0lrt}
was significant in the direction of $M_A$ at a one-tailed $\alpha=.05$, and
(iv) the $M_A$ BIC was lower than the $M_B$ BIC.  Of course, the fact that models are distinguishable does not necessarily imply that $M_A$
should be preferred to $M_B$.  However, the definitions above allow us
to put the tests on a common scale for the purpose of displaying results.

In addition to the above statistics, we computed two types of 90\% confidence
intervals of BIC differences (which are actually $\chi^2$ differences
here because the models have the same number of parameters).  The first
type of interval was based on
the result from~\eqref{eq:bicdiffint}, while the second type of
interval was based on the nonparametric bootstrap (based on 1,000
bootstrap samples per replication).  
Summaries of interest included
interval coverage, mean interval width, and interval variability.  The
latter statistic is defined as the pooled standard deviation of the lower and
upper confidence limits; for a given sample size and interval type, the statistic
is computed via:
\begin{equation}
    \label{eq:intvar}
    s_{\text{int}} = \sqrt{\frac{(n_{\text{rep}} - 1) \times (s^2_L +
        s^2_U)}{2 \times n_{\text{rep}} - 2}},
\end{equation}
where $s^2_L$ is the variance of the lower limit, $s^2_U$ is the
variance of the upper limit, and $n_{\text{rep}}$ is the number of
replications within one simulation condition (which is 3,000 for this
simulation).

\subsection{Results}

\begin{figure}
\caption{Power associated with test statistics, Simulation 1.}
\label{fig:sim1}
\begin{knitrout}\footnotesize
\definecolor{shadecolor}{rgb}{0.969, 0.969, 0.969}\color{fgcolor}

{\centering \includegraphics[width=6in,height=2in]{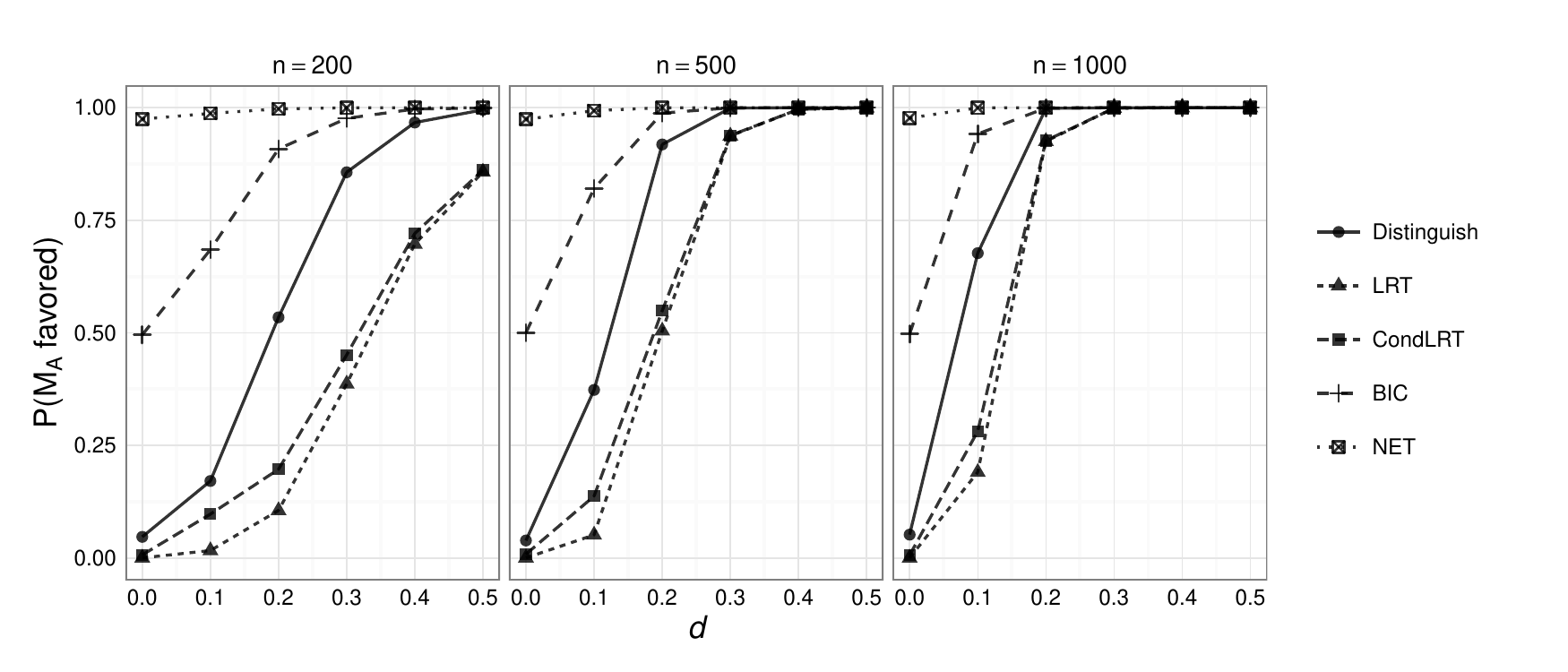} 

}

\end{knitrout}
\end{figure}

Overall simulation results are displayed in Figure~\ref{fig:sim1}.  The x-axes
display values of $d$, the y-axes display the probability that $M_A$
was preferred (using the criteria described previously), and panels display results for different
values of $n$.  The lines within each panel represent the four
statistics that were computed (with two lines for the LRT, further
described in the next paragraph).  We see that the NET procedure almost
never declares the two candidate models to be equivalent, even in the condition
where $d=0$.  This is because the NET is generally a test for global
equivalence, and the free paths that are unique to $M_A$ and $M_B$
result in model fits that are not equal to one another.
BIC, on the other hand, increasingly prefers the true
model, $M_A$, as $d$ and $n$ increase.  The problem, as
mentioned earlier, involves the fact that BIC provides no mechanism for
declaring models to be indistinguishable or to fit equally.  For example, in the
$(d=0.1, n=200)$ condition, BIC prefers $M_A$ about 70\% of the time.
The other 30\% of the time, BIC incorrectly prefers $M_B$.  In contrast, the Vuong tests provide a formal mechanism for concluding that neither model should be preferred (either because they are indistinguishable or because their fits are equal).

Focusing on the distinguishability test results in Figure~\ref{fig:sim1}, we
observe ``true'' Type I error rates in the $d=0$ condition: models are
incorrectly declared to be distinguishable approximately 5\% of the time.
Additionally, the hypothesis that models
are indistinguishable is increasingly rejected with both $d$ and $n$.
Finally, focusing on the Vuong LRT
results, we see near-zero Type I error rates in the $d=0$ conditions.  This
partially reflects the fact that the LRT should be used only when models are distinguishable (i.e., when $H_0$ for the distinguishability test is rejected).
To expand on this point, the ``CondLRT'' line displays power for the LRT conditioned on a
rejected $H_0$ for the distinguishability test.  These lines
show Type I error rates that are slightly closer to .05, though they are
still low.  This result matches the observation by \citeA{clamcc14}
that Vuong's sequential testing procedure can be conservative.  In particular,
\citeA{vuo89} proves an upper bound for the sequential tests' Type I
error that may actually be lower in practice.

Aside from Type I error, the power of the LRT
approaches 1 more slowly
 than the power of the distinguishability test.  The space between
the distinguishability and LRT curves (i.e., between the solid line and
the lower dashed lines of Figure~\ref{fig:sim1}) is related to the proportion of time
that the hypothesis of indistinguishability is rejected but the
hypothesis of equal-fitting models is not rejected.  This should not
be taken as evidence that the distinguishability test can be used
without the Vuong LRT.  If we reject the hypothesis of
distinguishability, we simply conclude that the models can potentially
be differentiated on the basis of fit.  We can draw no conclusions
about one model fitting better than the other.

Table~\ref{tab:sim1res} contains results associated with 90\%
confidence intervals of BIC differences.  The average interval width,
endpoint variability, and coverage are displayed in columns for both
the Vuong intervals and nonparametric bootstrap intervals.  Rows are
based on $n$ and $d$ (the value of the parameter labeled `A' in
Figure~\ref{fig:pathdiag1}).  It is seen that the two types of
interval estimates exhibit similar widths and endpoint variability.
Coverage is somewhat different, however.  When $d$
equals zero, models are indistinguishable and the intervals are
invalid.  This results in very high coverage rates across both types
of intervals.  As $d$ initially moves away from zero, both types
of intervals exhibit coverage that is too low.  Finally, as $d$
gets larger (and as $n$ increases), the intervals converge
towards the nominal coverage rate.  The bootstrap intervals have a
slight advantage here, moving towards a coverage of $0.9$ faster than
the Vuong intervals.

\begin{table}%
\caption{Average interval widths, variability in endpoints, and coverage of differences in non-nested BICs, Simulation 1.}
\label{tab:sim1res}
  \centering
\begin{tabular}{llcccccc}
\toprule
& & \multicolumn{2}{c}{Avg Width} & \multicolumn{2}{c}{Endpoint SD} & \multicolumn{2}{c}{Coverage} \\ \cmidrule(lr){3-4}\cmidrule(lr){5-6}\cmidrule(lr){7-8}
$n$ & Path A & Vuong & Boot & Vuong & Boot & Vuong & \multicolumn{1}{c}{Boot} \\ 
\midrule
200 & 0  & \multicolumn{1}{r}{$\phantom{0}8.351$} & \multicolumn{1}{r}{$10.907$} & \multicolumn{1}{r}{$\phantom{0}2.952$} & \multicolumn{1}{r}{$\phantom{0}3.159$} & \multicolumn{1}{r}{$\phantom{0}0.999$} & \multicolumn{1}{r}{$\phantom{0}0.995$} \\
 & 0.1  & \multicolumn{1}{r}{$11.174$} & \multicolumn{1}{r}{$13.051$} & \multicolumn{1}{r}{$\phantom{0}4.277$} & \multicolumn{1}{r}{$\phantom{0}4.496$} & \multicolumn{1}{r}{$\phantom{0}0.940$} & \multicolumn{1}{r}{$\phantom{0}0.979$} \\
 & 0.2  & \multicolumn{1}{r}{$17.052$} & \multicolumn{1}{r}{$17.779$} & \multicolumn{1}{r}{$\phantom{0}6.092$} & \multicolumn{1}{r}{$\phantom{0}6.113$} & \multicolumn{1}{r}{$\phantom{0}0.844$} & \multicolumn{1}{r}{$\phantom{0}0.883$} \\
 & 0.3  & \multicolumn{1}{r}{$23.631$} & \multicolumn{1}{r}{$23.359$} & \multicolumn{1}{r}{$\phantom{0}7.846$} & \multicolumn{1}{r}{$\phantom{0}7.526$} & \multicolumn{1}{r}{$\phantom{0}0.856$} & \multicolumn{1}{r}{$\phantom{0}0.884$} \\
 & 0.4  & \multicolumn{1}{r}{$29.427$} & \multicolumn{1}{r}{$28.452$} & \multicolumn{1}{r}{$\phantom{0}9.401$} & \multicolumn{1}{r}{$\phantom{0}8.867$} & \multicolumn{1}{r}{$\phantom{0}0.868$} & \multicolumn{1}{r}{$\phantom{0}0.885$} \\
 & 0.5  & \multicolumn{1}{r}{$33.607$} & \multicolumn{1}{r}{$32.772$} & \multicolumn{1}{r}{$10.731$} & \multicolumn{1}{r}{$10.352$} & \multicolumn{1}{r}{$\phantom{0}0.857$} & \multicolumn{1}{r}{$\phantom{0}0.874$} \\
500 & 0  & \multicolumn{1}{r}{$\phantom{0}8.256$} & \multicolumn{1}{r}{$10.697$} & \multicolumn{1}{r}{$\phantom{0}2.889$} & \multicolumn{1}{r}{$\phantom{0}3.127$} & \multicolumn{1}{r}{$\phantom{0}0.999$} & \multicolumn{1}{r}{$\phantom{0}0.995$} \\
 & 0.1  & \multicolumn{1}{r}{$14.460$} & \multicolumn{1}{r}{$15.762$} & \multicolumn{1}{r}{$\phantom{0}5.357$} & \multicolumn{1}{r}{$\phantom{0}5.692$} & \multicolumn{1}{r}{$\phantom{0}0.846$} & \multicolumn{1}{r}{$\phantom{0}0.896$} \\
 & 0.2  & \multicolumn{1}{r}{$25.645$} & \multicolumn{1}{r}{$25.939$} & \multicolumn{1}{r}{$\phantom{0}8.548$} & \multicolumn{1}{r}{$\phantom{0}8.607$} & \multicolumn{1}{r}{$\phantom{0}0.861$} & \multicolumn{1}{r}{$\phantom{0}0.888$} \\
 & 0.3  & \multicolumn{1}{r}{$37.184$} & \multicolumn{1}{r}{$36.880$} & \multicolumn{1}{r}{$11.690$} & \multicolumn{1}{r}{$11.438$} & \multicolumn{1}{r}{$\phantom{0}0.888$} & \multicolumn{1}{r}{$\phantom{0}0.899$} \\
 & 0.4  & \multicolumn{1}{r}{$47.055$} & \multicolumn{1}{r}{$45.960$} & \multicolumn{1}{r}{$14.536$} & \multicolumn{1}{r}{$13.906$} & \multicolumn{1}{r}{$\phantom{0}0.885$} & \multicolumn{1}{r}{$\phantom{0}0.895$} \\
 & 0.5  & \multicolumn{1}{r}{$53.824$} & \multicolumn{1}{r}{$52.641$} & \multicolumn{1}{r}{$16.581$} & \multicolumn{1}{r}{$16.004$} & \multicolumn{1}{r}{$\phantom{0}0.890$} & \multicolumn{1}{r}{$\phantom{0}0.896$} \\
1000 & 0  & \multicolumn{1}{r}{$\phantom{0}8.257$} & \multicolumn{1}{r}{$10.630$} & \multicolumn{1}{r}{$\phantom{0}2.960$} & \multicolumn{1}{r}{$\phantom{0}3.193$} & \multicolumn{1}{r}{$\phantom{0}0.999$} & \multicolumn{1}{r}{$\phantom{0}0.993$} \\
 & 0.1  & \multicolumn{1}{r}{$19.282$} & \multicolumn{1}{r}{$19.991$} & \multicolumn{1}{r}{$\phantom{0}6.701$} & \multicolumn{1}{r}{$\phantom{0}6.986$} & \multicolumn{1}{r}{$\phantom{0}0.860$} & \multicolumn{1}{r}{$\phantom{0}0.887$} \\
 & 0.2  & \multicolumn{1}{r}{$35.735$} & \multicolumn{1}{r}{$35.835$} & \multicolumn{1}{r}{$11.497$} & \multicolumn{1}{r}{$11.489$} & \multicolumn{1}{r}{$\phantom{0}0.876$} & \multicolumn{1}{r}{$\phantom{0}0.891$} \\
 & 0.3  & \multicolumn{1}{r}{$52.155$} & \multicolumn{1}{r}{$51.920$} & \multicolumn{1}{r}{$16.295$} & \multicolumn{1}{r}{$16.142$} & \multicolumn{1}{r}{$\phantom{0}0.895$} & \multicolumn{1}{r}{$\phantom{0}0.899$} \\
 & 0.4  & \multicolumn{1}{r}{$66.860$} & \multicolumn{1}{r}{$65.839$} & \multicolumn{1}{r}{$20.507$} & \multicolumn{1}{r}{$20.017$} & \multicolumn{1}{r}{$\phantom{0}0.893$} & \multicolumn{1}{r}{$\phantom{0}0.896$} \\
 & 0.5  & \multicolumn{1}{r}{$76.426$} & \multicolumn{1}{r}{$75.339$} & \multicolumn{1}{r}{$23.596$} & \multicolumn{1}{r}{$23.052$} & \multicolumn{1}{r}{$\phantom{0}0.893$} & \multicolumn{1}{r}{$\phantom{0}0.900$} \\
\bottomrule 
\end{tabular}

\end{table}

In practice, when models are more complex and generally easier to
distinguish, the Vuong intervals may exhibit coverage that is more
comparable to the bootstrap intervals regardless of $n$.
In the following simulation, we study this conjecture.

\section{Simulation 2: BIC Intervals}

The previous simulation showed that, when candidate models are nearly
indistinguishable from one another, interval estimates associated with
BIC differences (or with the likelihood ratio) generally stray from
the nominal coverage rate.  As models become more distinguishable, the intervals initially exhibit coverage
that is too high, followed by coverage that is too low, followed by
coverage that is just right.
In this
simulation, we further study the properties of these interval
estimates in more complex models that are generally distinguishable
from one another.  In this situation, the intervals' coverages should
be closer to their advertised coverages.

\subsection{Method}

The simulation was set up in a manner similar to the simulation from
\citeA{premer12}, using the previously-discussed models from Figure~\ref{fig:sim2}.
These models reflect four unique hypotheses about the relationships between 9
observed variables.
One thousand datasets were first generated from Model D, with
unstandardized path coefficients being fixed to 0.2, residual
variances being fixed to 0.8, and the exogenous variance associated
with variable $X_1$ being fixed to 1.0.
We then fit
Models A--C to each dataset and obtained interval estimates of
BIC differences.  We examined sample sizes of $n=200, 500,$
and 1,000 and
compared 90\% interval estimates from Vuong's theory to 90\% interval estimates
from the nonparametric bootstrap.  Statistics of interest were those
used in Simulation 1:
interval coverage, mean interval width, and interval variability.

\subsection{Results}

Results are displayed in Table~\ref{tab:sim2res}, with model pairs and
sample sizes in rows and interval statistics in
columns.   The two
columns on the right show that coverage is generally good for both
methods; the coverages are all close to $.9$.  The other columns show
that the Vuong intervals tend to be slightly better than the bootstrap
intervals: the Vuong widths are slightly smaller, and there is
slightly less variability in the endpoints.  These small advantages
may not be meaningful in many situations, but the results at least show
that the bootstrap intervals and Vuong intervals are comparable here.
The Vuong intervals have a clear computational
advantage, requiring only output from the two fitted models (and no
extra data sampling or model fitting).  As mentioned previously, the
bootstrap intervals may still exhibit better performance when
regularity conditions are violated.  In the following section, we
examine the Vuong statistics' application to nested model
comparison.

\begin{table}%
\caption{Average interval widths, variability in endpoints, and coverage of differences in non-nested BICs, Simulation 2.}
\label{tab:sim2res}
\centering
\begin{tabular}{llcccccc}
\toprule
& & \multicolumn{2}{c}{Avg Width} & \multicolumn{2}{c}{Endpoint SD} & \multicolumn{2}{c}{Coverage} \\ \cmidrule(lr){3-4}\cmidrule(lr){5-6}\cmidrule(lr){7-8}
Models & $n$ & Vuong & Boot & Vuong & Boot & Vuong & \multicolumn{1}{c}{Boot} \\ 
\midrule
A-B & 200  & $39.350$ & $42.241$ & $12.137$ & $12.521$ & $\phantom{0}0.919$ & $\phantom{0}0.873$ \\
 & 500  & $57.937$ & $59.620$ & $18.173$ & $18.516$ & $\phantom{0}0.901$ & $\phantom{0}0.875$ \\
 & 1000  & $79.614$ & $80.682$ & $23.484$ & $23.798$ & $\phantom{0}0.916$ & $\phantom{0}0.907$ \\
B-C & 200  & $45.469$ & $47.805$ & $14.508$ & $14.885$ & $\phantom{0}0.899$ & $\phantom{0}0.912$ \\
 & 500  & $69.064$ & $70.317$ & $20.832$ & $21.036$ & $\phantom{0}0.901$ & $\phantom{0}0.905$ \\
 & 1000  & $95.699$ & $96.403$ & $29.665$ & $29.953$ & $\phantom{0}0.893$ & $\phantom{0}0.891$ \\
C-A & 200  & $44.843$ & $48.082$ & $13.704$ & $14.107$ & $\phantom{0}0.919$ & $\phantom{0}0.894$ \\
 & 500  & $65.386$ & $67.320$ & $19.633$ & $19.853$ & $\phantom{0}0.910$ & $\phantom{0}0.891$ \\
 & 1000  & $89.381$ & $90.753$ & $27.655$ & $27.977$ & $\phantom{0}0.899$ & $\phantom{0}0.882$ \\
\bottomrule 
\end{tabular}

\end{table}

\section{Simulation 3: Tests of Nested Models}

When we first introduced the Vuong test statistics, 
we mentioned that the statistics $n\hat{\omega}^2_*$ and
$2n^{1/2}\text{LR}_{AB}$ provide unique tests of nested models.
Unlike the classical likelihood ratio test (also known as the $\chi^2$
difference test), tests involving these statistics make no assumption
related to the correctness of the full model.  In this simulation, we
compare the performance of the classical likelihood ratio test of
nested models to the Vuong tests involving $n\hat{\omega}^2_*$ and
$2n^{1/2}\text{LR}_{AB}$.

\subsection{Method}

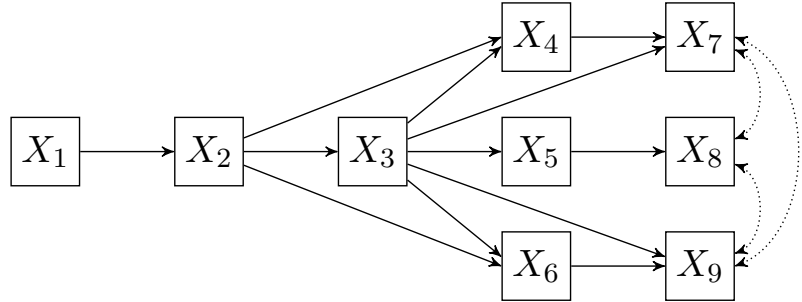
\begin{figure}%
  \caption{Path diagrams reflecting the models used in Simulation
    3.} \label{fig:sim3}
  \centering
  \resizebox{0.75\textwidth}{!}{
      \input{sim3}
    }
\end{figure}

The data generating model was the path model displayed in
Figure~\ref{fig:sim3}, where some parameters are represented by solid
lines, some parameters are represented by dashed lines, and variance
parameters are omitted from the figure.
Path coefficients (unstandardized) were set to $0.2$, residual
variances were set to $0.8$, and the single exogenous variance was set
to $1.0$.  The dashed covariance parameters, further described below, were
manipulated across conditions.
After generating data from
this model, we fit two candidate models to the data: the data
generating model ($M_A$), and a model with the dashed covariance paths
omitted ($M_B$).  The two candidate models are therefore nested, and
$M_A$ should be preferred when the dashed paths are nonzero.
This is a best-case scenario for the traditional likelihood ratio
test because the full model ($M_A$) is correct.

Simulation conditions were defined by $n$, which assumed values of 200, 500, and 1,000, and by the
value of the dashed covariance parameters, $d$.  Within a condition, these covariance
parameters simultaneously took values of $0, 0.025, 0.050, 0.075,
0.100$, or $0.125$.
In each condition,
we generated 1,000 datasets and fit both $M_A$ and $M_B$ to the
data.  We then computed four statistics: the distinguishability test
of~\eqref{eq:h0om2} (which is actually a test of nested models here), the Vuong LRT involving $2n^{1/2}\text{LR}_{AB}$,
the classical LRT based on $\chi^2$ differences,
and the BIC difference.  For each statistic, we recorded whether or
not $M_A$ was favored over $M_B$ using the same criteria that were
used in Simulation 1.

\subsection{Results}

\begin{figure}
\caption{Power associated with test statistics, Simulation 3.}
\label{fig:sim3res}
\begin{knitrout}\footnotesize
\definecolor{shadecolor}{rgb}{0.969, 0.969, 0.969}\color{fgcolor}

{\centering \includegraphics[width=6in,height=2in]{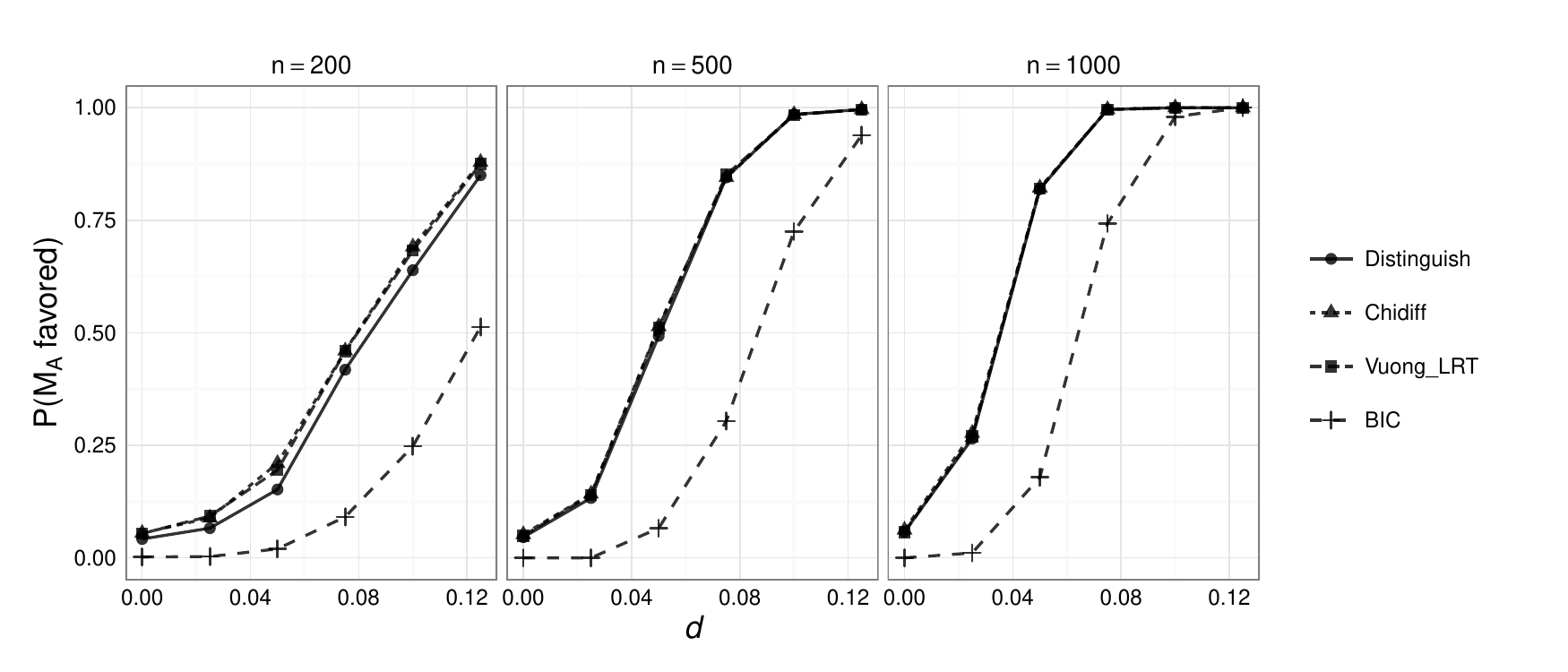} 

}

\end{knitrout}
\end{figure}

Results are displayed in Figure~\ref{fig:sim3res}, which is arranged
similarly to the figure from Simulation 1.  It is seen that power
associated with the two Vuong tests is very similar to power
associated with the classical LRT (labeled ``Chidiff'' in the figure), with the distinguishability test exhibiting slightly smaller power at $n=200$.  In the $n=500$ and $n=$1,000 conditions, the two Vuong statistics are nearly equivalent to the classical LRT.
The BIC, on the other hand,
is slower to prefer $M_A$ due to the fact that $M_A$ has three extra
parameters.  These extra parameters must be sufficiently far from zero
before they are ``worth it,'' as judged by BIC.

This simulation demonstrates that, within the controlled environment examined here,
little is lost if one uses Vuong tests to compare
nested models.  In some conditions, the Vuong statistics even exhibited
slightly higher power than the classical LRT.  \citeA{vuo89} showed
that the limiting distributions of his statistics matched the
distribution of the classical LRT (when the full model is correctly
specified), and our simulation provides
evidence that his statistics approach the limiting distribution at a
similar rate as the classical LRT.  This is relevant to sample size
considerations: the Vuong statistics exhibited similar power
to the classical LRT, so that we can potentially use classical LRT
results to guide us on sample sizes that lead to sufficient power for
the Vuong tests.  Simulation studies can also be carried out to
study the power of the Vuong tests in specific situations.

Future
work could compare the Vuong statistics to both the classical LRT and
to other robust statistics \cite<see>{satben94,sav14} when the candidate models are both incorrect.
This is trickier than it sounds, because it is difficult to design a
framework that allows us to assess the statistics' Type I error
rates.  Such a framework would require that (i) the null hypothesis
from~\eqref{eq:tradh0} holds, and
(ii) the data generating model differs from the full model.  This
framework is necessary so that we know whether or not differences in
statistics' power are paired with differences in statistics' Type I
error rates.

\section{General Discussion}

The framework described in this paper provides researchers with a
general means to test pairs of SEMs for differences in fit.
Researchers also gain a means to test whether pairs of SEMs
are distinguishable in a population of interest
and to test
for differences between models' information criteria.  In the
discussion below, we provide detail on extension of
the tests to comparing multiple ($>2$) models.  We
also address regularity conditions, extension to other types of models, and a recommended strategy for
applied researchers.

\paragraph{Comparing Multiple Models}

The reader may wonder whether or not the above theory extends
to simultaneously testing multiple models.  \citeA{kat08} derives the joint
distribution of LR statistics comparing $(m-1)$ models to a baseline
model (i.e., when we have a unique LR statistic comparing each of
$(m-1)$ models to the baseline model) and obtains a
test statistic based on the sum of the $(m-1)$ squared LR statistics.
We do not describe all the details here but instead supply some
informal intuition underlying the tests.

In a situation where one wishes to compare multiple models, we can
obtain the casewise likelihoods
$\ell_j(\bm{\theta}; x_i)$, $i=1,\ldots,n$, $j=1,\ldots,m$.
We could then subject these casewise likelihoods
to an ANOVA, where case ($i$) is a between-subjects factor
and model ($j$) is a within-subjects factor.  We should
expect a main effect of case, because the models will naturally fit
some individuals better than others.  The main effect of model,
however, serves as a test of whether or not all $m$ model fits are
equal.  Additionally, the error variance informs us of the extent to
which models are distinguishable: if the error variance is close to zero, then we
have evidence for indistinguishable models.  The ANOVA framework could also
be useful for posthoc tests, whereby one wishes to specifically know
which model fits differ from which others.

We have not implemented the tests derived by Katayama, and the
ANOVA described above is not the same as Katayama's tests.  For
example, the ANOVA assumes sphericity for the
within-subjects ``model'' factor, whereas Katayama explicitly estimates
covariances between model likelihoods.  Simultaneous tests of multiple
SEMs generally provide interesting directions for future research.

\paragraph{Regularity Conditions} We previously outlined the regularity
conditions associated with the Vuong tests, and we expand here on the conditions'
relevance to applied researchers.  The conditions under which the
Vuong tests hold are fairly general, including existence of
second-order derivatives of the log-likelihood, invertibility of the models'
information matrices, and independence and identical distribution of
the data vectors $\bm{x}_i$.  The invertibility requirement can be
violated when we apply the distinguishability test to compare mixture models
with different numbers of components.  This violation is related to the fact
that a model with $m-1$ components lies on the boundary of the
parameter space of a model with $m$ components.  \citeA{jef03} presents some
evidence that this can
cause inflated Type I error rates, and \citeA{wil15} further describes
problems with the test's application to mixture models.  At the time
of this writing, researchers have generally ignored these issues.
Similar issues may arise if one model can be obtained from the other by setting some variance parameters equal to zero.
The iid requirement can be violated by
certain time series models.  This implies that the tests described
here may not be applicable to some dynamic SEMs (where $n$ reflects
observed timepoints), though \citeA{rivvuo02} extend the
tests to handle these types of models.

\paragraph{Recommended Use}
Using the NET and the methods described here \cite<and also
in>{levhan07,levhan11}, one gains a fuller comparison of candidate models.
These procedures give researchers the ability to routinely test
non-nested models for global equivalence, distinguishability,
and differences in fit or information criteria.
We recommend the following non-nested model comparison sequence.\\ \ \\
\begin{enumerate}
  \item Using the NET, evaluate models for global equivalence
    (can be done prior to data collection).  If models are not
    found to be equivalent, proceed to 2.
  \item Test whether or not models are distinguishable, using the $\omega^2_*$
    statistic (data must have already been collected).  If models are
     found to be distinguishable, proceed to 3.
  \item Compare models via the non-nested LRT or a confidence
    interval of BIC differences.\\ \ \\
\end{enumerate}
If one makes it to the third step, then the test or interval
estimate may allow for the preference of one model.  Otherwise, we
cannot prefer either model to the other.
In the case that models are found to be indistinguishable or to have
equal fit, follow-up modeling can often be performed to further
study the data.  For example, one can often specify a larger model
that encompasses both candidate models as special cases.  This can
provide information about important model parameters, which may lead to an alternative model that is a cross between the
original candidate models.  Overfitting would be a concern associated
with this strategy, and it may often be sufficient to simply report
results of the encompassing model.  Aside from this strategy, however,
there is nothing inherently wrong with having indistinguishable or
equal-fitting models.  Sometimes, there is simply not enough
information in the data to differentiate two theories.
The above suggested sequence provides more information about the
models' relative standings than do traditional comparisons via BIC,
which should help researchers to favor a model only when the data
truly favor that model.

\paragraph{Distributional Assumptions}
While our current implementation allows researchers to carry out the
above steps using SEMs estimated under multivariate normality,
extensions to other assumed distributions is immediate.  For example,
if our observed variables are ordinal, our model may be based on a
multinomial distribution.  If the model is estimated via ML, then the
tests can be carried out as described here; we just need to obtain the
``non-standard'' model output (casewise contributions to the
likelihood, scores) for those models.  Researchers often elect to
employ alternative discrepancy functions that do not correspond to a
specific probability distribution, however, especially when the
observed data are discrete.  Examples of estimation methods that do
not rely on well-defined probability distributions include weighted
least squares and pairwise maximum likelihood \cite<e.g.,>{katmou12,mut84}.
Work by
\citeA{gol03} implies that Vuong's theory can also be applied to
models estimated via these alternative discrepancy functions.  Further
work is needed to obtain the necessary output from models estimated
under these functions and to study the tests' applications.


Finally, we note that the ideas described throughout this paper
generally apply to situations where one's goal is to declare a single
model as the best.  One may instead wish to
average over the set of candidate models, drawing general conclusions
across the set \cite<e.g.,>{hoemad99}.  Though it is computationally
more difficult, the
model averaging strategy allows the researcher to explicitly
acknowledge that all of the models in the set are ultimately
incorrect.

\section*{Computational Details}

All results were obtained using the \proglang{R}~system for statistical computing \cite{R11},
version~3.2.0, employing the add-on packages
\pkg{lavaan}~0.5-18
\cite{lavaan11} for fitting of the models and score computation,
\pkg{nonnest2}~0.2
\cite{nonnest2} for carrying out the Vuong tests, and
\pkg{simsem}~0.5-8
\cite{simsempack} for simulation convenience functions.
\proglang{R}~and the packages \pkg{lavaan}, \pkg{nonnest2}, and \pkg{simsem} are
freely available under the General Public License~2 from the
Comprehensive \proglang{R} Archive Network at \url{http://CRAN.R-project.org/}.
\proglang{R}~code for replication of our results is available at
\url{http://semtools.R-Forge.R-project.org/}.

\bibliography{refs}

\appendix

\section{Tests involving weighted sums of $\chi^2$ distributions}
\label{app:var}

In this appendix, we describe technical details underlying the tests
whose limiting distributions are weighted sums of $\chi^2$ statistics.
As stated in the main text, under
the hypothesis that $\omega^2_*=0$, $n \hat{\omega}^2_*$ converges in
distribution to a weighted sum of $(k+q)$ chi-square distributions (with
1 degree of freedom each),
where $k$ and $q$ are the number of free parameters in $M_A$ and
$M_B$, respectively.  The weights are defined as the squared
eigenvalues of a matrix $\bm{W}$, which is defined below.
Additionally, for nested or indistinguishable models, $2n^{1/2}\text{LR}_{AB}$
converges in distribution to a similar weighted sum of chi-square
distributions, with weights defined as the {\em unsquared} eigenvalues
of the same matrix $\bm{W}$.

To obtain $\bm{W}$,
let the matrices $\bm{U}_A(\bm{\theta}_A)$ and
$\bm{V}_A(\bm{\theta}_A)$ be defined as
\begin{align*}
    \bm{U}_A(\bm{\theta}_A) &= E \left [ \frac{\partial^2
          \ell(\bm{\theta}_A; \bm{x}_i)}{\partial
          \bm{\theta}_A \partial \bm{\theta}_A^\prime} \right ] \\
    \bm{V}_A(\bm{\theta}_A) &= E \left [ \frac{\partial
          \ell(\bm{\theta}_A; \bm{x}_i)}{\partial
          \bm{\theta}_A} \cdot \frac{ \partial
          \ell(\bm{\theta}_A; \bm{x}_i)}{\partial \bm{\theta}_A^\prime}
    \right ],
\end{align*}
which can be obtained from a fitted $M_A$'s information matrix and
cross-product of scores (see Equations~\eqref{eq:score} and~\eqref{eq:infmat}),
respectively.  The matrices $\bm{U}_B(\bm{\theta}_B)$ and
$\bm{V}_B(\bm{\theta}_B)$ are defined similarly.  
Further, define
$\bm{V}_{AB}(\bm{\theta}_A, \bm{\theta}_B)$ as
\begin{equation*}
    \label{eq:vab}
    \bm{V}_{AB}(\bm{\theta}_A, \bm{\theta}_B) = E \left [ \frac{\partial
          \ell(\bm{\theta}_A; \bm{x}_i)}{\partial
          \bm{\theta}_A} \cdot \frac{ \partial
          \ell(\bm{\theta}_B; \bm{x}_i)}{\partial \bm{\theta}_B^\prime}
    \right ],
\end{equation*}
which can be obtained by taking products of ${\bm s}(\hat{{\bm
  \theta}}_A; \bm{x}_i)$ and ${\bm s}(\hat{{\bm \theta}}_B; \bm{x}_i)$.
The matrix $\bm{W}$ is then defined as
\begin{equation*}
    \label{eq:wmat}
    \bm{W} = \left [ \begin{array}{cc}
            -\bm{V}_A(\bm{\theta}_A) \bm{U}^{-1}_A(\bm{\theta}_A) &
            -\bm{V}_{AB}(\bm{\theta}_A, \bm{\theta}_B)
            \bm{U}^{-1}_B(\bm{\theta}_B) \\
            \bm{V}^\prime_{AB}(\bm{\theta}_A, \bm{\theta}_B)
            \bm{U}^{-1}_A(\bm{\theta}_A) &
            \bm{V}_B(\bm{\theta}_B) \bm{U}^{-1}_B(\bm{\theta}_B)
            \end{array} \right ].
\end{equation*}
As stated previously, the eigenvalues of $\bm{W}$ then determine the weights involved in the limiting sum of $\chi^2$ distributions.  See \citeA{vuo89} for the proof and further detail.



\end{document}

%% file: simDiagram.tex
\begin{tikzpicture}
	\matrix[row sep=1.5mm,ampersand replacement=\&] {
		\node(X1)[manifest]{$X_{1}$}; \& \&[1em]
		\node(X2)[manifest]{$X_{2}$}; \& \&[1em]
		\node(X3)[manifest]{$X_{3}$}; \& \&[1em]
		\node(X4)[manifest]{$X_{4}$}; \& \&[1em]
		\node(X5)[manifest]{$X_{5}$}; \& \&[1em]
		\node(X6)[manifest]{$X_{6}$}; \\
	};

	\node(F1)[latent] at ($(X2) + (0em,7em)$) {$F_1$};
	\node(F2)[latent] at ($(X5) + (0em,7em)$) {$F_2$};

\foreach \n in {1,2,3}{
	\draw[-stealth'] (F1)--(X\n.north);
}

\foreach \n in {4,5,6}{
	\draw[-stealth'] (F2)--(X\n.north);
}

\foreach \n in {1,2,3,4,5,6}{
	\node(E\n)[error] at ($(X\n) + (0em,-3.5em)$) {$\scriptstyle E_\n$};
        \draw[-stealth'] (E\n.north)--(X\n.south);
}

	\draw[-stealth',gray,densely dashdotted] (F1)--(X4.north) node[rectangle,pos=0.4,fill=white,inner sep=0.3ex] {$\scriptstyle A$};
	\draw[-stealth',gray,densely dashdotted] (F2)--(X3.north)
        node[rectangle,pos=0.4,fill=white,inner sep=0.3ex]
        {$\scriptstyle B$};
        \draw[stealth'-stealth'] (F1.north) to [out=70, in=110] (F2.north);

\end{tikzpicture}

%% file: modelA.tex
\begin{tikzpicture}
  [manifest/.style={rectangle,draw,inner sep=0.15em,minimum size=1.8em}]
  \matrix[row sep={10ex,between origins}, column sep={7ex,between origins}, ampersand replacement=\&]
  {
    \& \&
    \node(X1)[manifest]{$X_{1}$}; \&
    \node(X2)[manifest]{$X_{2}$}; \& \&  \\
    \& \& \& \& \& \\
    \node(X4)[manifest]{$X_{4}$}; \&
    \node(X5)[manifest]{$X_{5}$}; \&
    \node(X6)[manifest]{$X_{6}$}; \&
    \node(X7)[manifest]{$X_{7}$}; \&
    \node(X8)[manifest]{$X_{8}$}; \&
    \node(X9)[manifest]{$X_{9}$}; \\
  };
  \node(X3)[manifest] at ($(X1) + (3.5ex,-10ex)$) {$X_{3}$};
  \foreach \n in {1,2}{
    \draw[-stealth'] (X\n.south) -- (X3);
  }
  \foreach \n in {4,5,6,7,8,9}{
    \draw[-stealth'] (X3) -- (X\n.north);
  }

  \draw[stealth'-stealth'] (X1.north) to [out=75, in=105] (X2.north);
\end{tikzpicture}

%% file: modelB.tex
\begin{tikzpicture}
  [manifest/.style={rectangle,draw,inner sep=0.15em,minimum size=1.8em}]
  \matrix[row sep={7ex,between origins}, column sep={10ex,between origins}, ampersand replacement=\&]
  {
    \& \& \&
    \node(X4)[manifest]{$X_{4}$}; \&
    \node(X7)[manifest]{$X_{7}$}; \\
    \node(X1)[manifest]{$X_{1}$}; \&
    \node(X2)[manifest]{$X_{2}$}; \&
    \node(X3)[manifest]{$X_{3}$}; \&
    \node(X5)[manifest]{$X_{5}$}; \&
    \node(X8)[manifest]{$X_{8}$}; \\
    \& \& \&
    \node(X6)[manifest]{$X_{6}$}; \&
    \node(X9)[manifest]{$X_{9}$}; \\
  };

  \draw[-stealth'] (X1) -- (X2.west);

  \foreach \n in {3,4,6}{
    \draw[-stealth'] (X2) -- (X\n.west);
  }

  \draw[-stealth'] (X3) -- (X5.west);
  \foreach \n in {4,7}{
    \draw[-stealth'] (X3) -- (X\n.195);
  }
  \foreach \n in {6,9}{
    \draw[-stealth'] (X3) -- (X\n.165);
  }

  \draw[-stealth'] (X4) -- (X7.west);

  \draw[-stealth'] (X5) -- (X8.west);

  \draw[-stealth'] (X6) -- (X9.west);

  \draw[stealth'-stealth'] (X7.east) to [out=340, in=20] (X9.east);
  \draw[stealth'-stealth'] (X7) to [out=340, in=20] (X8);
  \draw[stealth'-stealth'] (X8) to [out=340, in=20] (X9);
\end{tikzpicture}

%% file: modelC.tex
\begin{tikzpicture}
  [manifest/.style={rectangle,draw,inner sep=0.15em,minimum size=1.8em}]
  \matrix[row sep={10ex,between origins}, column sep={7ex,between origins}, ampersand replacement=\&]
  {
    \node(X1)[manifest]{$X_{1}$}; \&
    \node(X2)[manifest]{$X_{2}$}; \&
    \node(X3)[manifest]{$X_{3}$}; \&
    \node(X4)[manifest]{$X_{4}$}; \&
    \node(X5)[manifest]{$X_{5}$}; \& \\
    \&
    \node(X6)[manifest]{$X_{6}$}; \&
    \node(X7)[manifest]{$X_{7}$}; \&
    \node(X8)[manifest]{$X_{8}$}; \&
    \& \\
    \& \&
    \node(X9)[manifest]{$X_{9}$}; \&
    \&  \\
  };

  \foreach \from/\to in {3/6, 4/7, 5/8}{
    \draw[-stealth'] (X\from.south) -- (X\to.north);
  }

  \foreach \from/\to in {6/9, 7/9, 8/9}{
    \draw[-stealth'] (X\from.south) -- (X\to);
  }

  \foreach \from/\to/\df/\dt in {1/2/60/120, 2/3/60/120, 3/4/60/120,
    4/5/60/120, 1/3/75/105, 2/4/75/105, 3/5/75/105, 1/4/90/90,
    2/5/90/90, 1/5/105/75}{
    \draw[stealth'-stealth'] (X\from.\df) to [out=45, in=135] (X\to.\dt);
  }
\end{tikzpicture}

%% file: modelD.tex
\begin{tikzpicture}
  [manifest/.style={rectangle,draw,inner sep=0.15em,minimum size=1.8em}]
  \matrix[row sep={10ex,between origins}, column sep={7ex,between origins}, ampersand replacement=\&]
  {
    \& \&
    \node(X1)[manifest]{$X_{1}$}; \&
    \&  \\
    \&
    \node(X2)[manifest]{$X_{2}$}; \&
    \node(X3)[manifest]{$X_{3}$}; \&
    \node(X4)[manifest]{$X_{4}$}; \&
    \& \\
    \node(X5)[manifest]{$X_{5}$}; \&
    \node(X6)[manifest]{$X_{6}$}; \&
    \node(X7)[manifest]{$X_{7}$}; \&
    \node(X8)[manifest]{$X_{8}$}; \&
    \node(X9)[manifest]{$X_{9}$}; \& \\
  };

  \foreach \from/\to in {1/2, 1/3, 1/4, 2/5, 2/6, 2/7, 3/7, 3/8, 4/8, 4/9}{
    \draw[-stealth'] (X\from) -- (X\to);
  }

\end{tikzpicture}

%% file: burnoutDiagram.tex
\begin{tikzpicture}
  [box/.style={circle,draw,inner sep=0.3ex,minimum size=3.5ex}]

  \node (PA) at (7.7, -0.3) [box] {PA};

  \node (RA) at (7.7, 1.4) [box] {RA}
    edge [-stealth'] (PA);

  \node (DP) at (5.2, 2.6) [box] {DP}
    edge [-stealth'] (PA);

  \node (EE) at (5.2, 0.5) [box] {EE}
    edge [-stealth'] (DP)
    edge [-stealth',densely dotted] node[rectangle,pos=0.4,fill=white,inner sep=0.3ex] {$\scriptstyle -M_3$} (PA);

  \node (ELC) at (5.6,-2.5) [box] {ELC}
    edge [-stealth',densely dotted] node[rectangle,pos=0.5,fill=white,inner sep=0.3ex] {$\scriptstyle -M_2$} (PA);

  \node (CC) at (2.7, 2) [box] {CC}
    edge [-stealth'] (DP)
    edge [-stealth'] (EE);

  \node (RC) at (2.7, 1) [box] {RC}
    edge [-stealth'] (EE)
    edge [-stealth',densely dotted] node[rectangle,pos=0.5,fill=white,inner sep=0.3ex] {$\scriptstyle -M_2$} (DP)
    edge [-stealth',densely dotted] node[rectangle,pos=0.6,fill=white,inner sep=0.3ex] {$\scriptstyle +M_2,M_3$} (ELC);

  \node (SE) at (0, -0.3) [box] {SE}
    edge [-stealth'] (EE)
    edge [-stealth'] (PA)
    edge [-stealth'] (ELC);

  \node (DM) at (1.4,-2) [box] {DM}
    edge [-stealth'] (SE)
    edge [-stealth',densely dotted] node[rectangle,pos=0.3,fill=white,inner sep=0.3ex] {$\scriptstyle -M_2$} (ELC);

  \node (SS) at (1, 1) [box] {SS}
    edge [-stealth'] (SE);

  \node (PS) at (2.8,-0.8) [box] {PS}
    edge [-stealth',densely dotted] node[rectangle,pos=0.4,fill=white,inner sep=0.3ex] {$\scriptstyle -M_2$} (SE);
    \draw [-stealth', densely dotted] (PS) to node[rectangle,pos=0.3,fill=white,inner sep=0.3ex] {$\scriptstyle +M_2$} (ELC);
\end{tikzpicture}

%% file: sim3.tex
\begin{tikzpicture}
  [manifest/.style={rectangle,draw,inner sep=0.15em,minimum size=1.8em}]
  \matrix[row sep={7ex,between origins}, column sep={10ex,between origins}, ampersand replacement=\&]
  {
    \& \& \&
    \node(X4)[manifest]{$X_{4}$}; \&
    \node(X7)[manifest]{$X_{7}$}; \\
    \node(X1)[manifest]{$X_{1}$}; \&
    \node(X2)[manifest]{$X_{2}$}; \&
    \node(X3)[manifest]{$X_{3}$}; \&
    \node(X5)[manifest]{$X_{5}$}; \&
    \node(X8)[manifest]{$X_{8}$}; \\
    \& \& \&
    \node(X6)[manifest]{$X_{6}$}; \&
    \node(X9)[manifest]{$X_{9}$}; \\
  };

  \draw[-stealth'] (X1) -- (X2.west);

  \foreach \n in {3,4,6}{
    \draw[-stealth'] (X2) -- (X\n.west);
  }

  \draw[-stealth'] (X3) -- (X5.west);
  \foreach \n in {4,7}{
    \draw[-stealth'] (X3) -- (X\n.195);
  }
  \foreach \n in {6,9}{
    \draw[-stealth'] (X3) -- (X\n.165);
  }

  \draw[-stealth'] (X4) -- (X7.west);

  \draw[-stealth'] (X5) -- (X8.west);

  \draw[-stealth'] (X6) -- (X9.west);

  \draw[stealth'-stealth',densely dotted] (X7.east) to [out=340, in=20] (X9.east);
  \draw[stealth'-stealth',densely dotted] (X7) to [out=340, in=20] (X8);
  \draw[stealth'-stealth',densely dotted] (X8) to [out=340, in=20] (X9);
\end{tikzpicture}